\documentclass[%
aip,
cp,amsmath,amssymb,
reprint,%
]{revtex4-2}
\usepackage{graphicx}
\usepackage{dcolumn}
\usepackage{bm}

\usepackage[utf8]{inputenc}
\usepackage[T1]{fontenc}
\usepackage{mathptmx,physics}
\usepackage{etoolbox}

\makeatletter
\def\@email#1#2{%
	\endgroup
	\patchcmd{\titleblock@produce}
	{\frontmatter@RRAPformat}
	{\frontmatter@RRAPformat{\produce@RRAP{*#1\href{mailto:#2}{#2}}}\frontmatter@RRAPformat}
	{}{}
}%
\makeatother
\begin{document}
	\preprint{AIP/123-QED}
	
	\title{Nonclassical Properties of a State Generated by a Driven Dispersive Interaction}
	\author{Naveen Kumar$^1$}
	\email{$^1$naveen74418@gmail.com and $^2$arpita.sps@gmail.com}
	\author{Arpita Chatterjee$^2$}
	\affiliation{Department of Mathematics,J. C. Bose University of Science and Technology, YMCA\text{,} Faridabad-121006\text{,} Haryana\text{,} India}
	\begin{abstract}
		We consider a cavity field state, which is created by the atom-cavity field's interaction in the presence of a driven field. The two-level atom passes through the cavity and is driven by a weak classical field. A photon number dependent Stark shift is induced by the atom's dispersive interaction with the cavity field. When the atom is in excited state $|a\rangle$, the output cavity field thus obtained is taken into consideration. With the help of the state evaluated, we investigate different statistical properties such as photon number distribution, Wigner function, Mandel's $Q$ parameter and squeezing.
	\end{abstract}
	
	\maketitle
	
	\section{\label{sec:level1}INTRODUCTION}

	The production and manipulation of nonclassical light fields are of great interest in quantum optics and quantum information processing \cite{bouwmeester00}. Nonclassical states have many real life applications, for example, squeezed states are used to reduce the noise level in one of the phase-space quadratures
below the quantum limit \cite{scully97}, entangled states produced
in down-conversion process are employed to test fundamental quantum features such as non-locality \cite{zeilinger98} and to
realize quantum information transmission schemes (cryptography \cite{braunstein05,naik00} or teleportation \cite{bouwmeester98}). Quantum superpositions of fields with different classical parameters are used to explore the quantum or classical boundary and the decoherence phenomenon \cite{brune96}.

In this context, theoreticians and experimentalists have offered many different strategies to produce nonclassical states of the optical field. To produce a nonclassical state adding and/or subtracting photons to standard quantum states is the right approach. The coherent states after adding a number of photons exhibit the gradual shifting from spontaneous to stimulated regimes of light emission \cite{zavatta04}. Also, subtraction of photons can be used to improve the Gaussian state's entanglement \cite{ourjoumtsev07, browne03} and quantum computing \cite{bartlett02}.
	
A single-photon Fock state is an essential resource in quantum information processing devices \cite{knill01}. These one-photon Fock states can be prepared by using cavity QED experiments in which atoms interact one at a time with a high $Q$ resonator, a quantum $\pi$ Rabi pulse in a microwave cavity  \cite{maitre97} or by an adiabatic passage sequence of atoms in an optical cavity \cite{henrich00}. We report here how an interacting atom-cavity field system in presence of a weak classical field generates a nonclassical field.

In this article, we describe the nonclassical properties of the cavity field state. We consider the interaction between a two-level atom and a single-mode cavity field driven by a weak classical field. To find the desired state of interest we solve the time-dependent Schrodinger equation with the effective Hamiltonian of the atom-cavity field system. To check the nonclassical nature of the evaluated state, we discuss some properties like the Wigner function and Mandel's $Q$ parameter. The negativity of the Wigner function affirms the nonclassicality of the state. The study of these states provides an essential understanding of quantum communication and quantum teleportation technique.

	\section{State of interest}

We consider a two-level atom interacting with a singlemode cavity field and driven by a weak classical field.
The Hamiltonian (assuming $\hbar = 1$) describing the atom-field interaction in the rotating-wave approximation is
\cite{alsing,biao,arpita}
	\begin{eqnarray}\nonumber
		H = \omega_0 S_z + \omega_a a^\dag a + g(a^\dag S^- + aS^+)
		+
		\epsilon (S^+ e^{-i\omega t} + S^- e^{i\omega t})\\
	\end{eqnarray}
where all the notations have their usual meanings (cf. Table~\ref{t1}).
If the system's reference frame is rotated with respect to the driving field frequency $\omega$, the Hamiltonian is given by
	$\omega$
	\begin{equation}
		H_f = \Delta S_z+\delta a^\dag a+g(a^\dag
		S^-+aS^+)+\epsilon(S^++S^-)\\
	\end{equation}
	where $\Delta = \omega_0-\omega$, $\delta = \omega_a-\omega$.
	For further simplification, considering  $\Delta=0$, the interaction Hamiltonian is
	\begin{eqnarray*}
		H_I & = & e^{iH_{f_0}t}H_{f_1} e^{-iH_{f_0}t}\\
		& = & \frac{1}{2}g\left[|+\rangle\langle+|-|-\rangle\langle-|+e^{i2\epsilon t}|+\rangle\langle-|
		- e^{-i2\epsilon t}|-\rangle\langle+|\right]\\
		& & \times ae^{-i\delta t}+\mathrm{H.C.}
	\end{eqnarray*}
	where
	\begin{eqnarray*}
		H_{f_0} = \delta a^\dag a+\epsilon(S^++S^-)
	\end{eqnarray*}
	and
	\begin{eqnarray*}
		H_{f_1} = g(a^\dag S^-+aS^\dag)
	\end{eqnarray*}
	and the dressed states $|\pm\rangle = \frac{1}{\sqrt{2}}(|a\rangle\pm|b\rangle)$
	are the eigenstates of $(S^++S^-)$ with the eigenvalues $\pm 1$.
	In the strong driving regime, $\epsilon\gg\{\delta, g\}$. As a result, high-frequency oscillating terms can be discarded. Now our effective Hamiltonian $H_{\mathrm{eff}}$ is \cite{chen}
	\begin{eqnarray}\nonumber
		H_{\mathrm{eff}} =\frac{1}{2}g (aS^+e^{-i\delta t}+a^\dag S^- e^{i\delta t})
	\end{eqnarray}
	
	With help of this effective Hamiltonian $H_{\mathrm{eff}}$, the time-dependent Schrodinger equation  \cite{scully97}
	\begin{eqnarray}
		\frac{i\partial | \psi(t) \rangle }{\partial t}= H_{\mathrm{eff}} |\psi (t)\rangle,
	\end{eqnarray}
	is now solved for an arbitrary state vector $|\psi(t)\rangle = \sum {c_{a, n}(t) |a, n\rangle +c_{b, n+1}(t) |b, n+1\rangle}(t)$ with probability amplitudes $c_{a,n}(t)$
	and $c_{b,n+1}(t)$, $|a\rangle (|b\rangle)$ is the excited (ground) state. The equations of motion in terms of $\dot{c}_{a,n}(t)$ and $\dot{c}_{b,n+1}(t)$ are
	\begin{eqnarray*}
		\dot{c}_{a,n}(t) & = & -\frac{ig\sqrt{n+1}}{2}  e^{-i\delta t} c_{b,n+1}(t)\\
		\dot{c}_{b,n+1}(t) & = & -\frac{ig\sqrt{n+1}}{2}  e^{i\delta t} c_{a,n}(t)
	\end{eqnarray*}
	Initially if the atom enters the cavity in the excited state $|a \rangle$ then $c_{a,n}(0) = c_n(0)$ and $c_{b,n+1}(0) = 0$, where $c_n(0)$ is the probability amplitude for the field
alone. A general solution with these conditions is
	\begin{eqnarray*}
		c_{a,n} & = & c_{n}(0)\left\lbrace \cos{\frac{\Omega_{n}t}{2}}-\frac{i\delta}{\Omega_{n}} \sin{\frac{\Omega_{n}t}{2}}\right\rbrace e^{\frac{i\delta t}{2}},\\
		c_{b,n+1} & = & -c_{n}(0)  \frac{2ig\sqrt{n+1}}{\Omega_{n}} \sin{\frac{\Omega_{n}t}{2}} e^{\frac{i\delta t}{2}},
	\end{eqnarray*}
	where ${\Omega_n}^2=\delta^2 +4g^2(n+1)$. If the radiation field is initially in a coherent state then $c_{n}(0)=e^{-\frac{|\alpha|^2}{2}}\frac{\alpha^n}{\sqrt{n!}}$ \cite{peng1}. The state vector $|\psi(t)\rangle$ describes the time evolution of the entire system of atom and cavity field but we now concentrate on some statistical aspects of the cavity field only. The field inside the cavity after departing the atom is
obtained by tracing out the atomic part of $\rho(t)$ as
	\begin{equation}
		\label{eq10}
		{\rho(t)}_{f} = Tr_{a}[\rho(t)],
	\end{equation}
where we have used the subscript $a (f)$ to denote the
atom (field).
	This ${\rho(t)}_{f}$ will be of consideration in all of the further work to determine the statistical properties of the field left into the cavity.
	\begin{table}
		\caption{Symbols used \label{t1}}
		\begin{ruledtabular}
			\begin{tabular}{lcr}
				&  &\\
				\hline
				$a^\dag$ & creation operator \\
				\hline
				$a$ &   annihilation operator  \\
				\hline
				$S_z$ &  inversion operator for the atom   \\
				\hline
				$S^+$ &  raising operator for the atom \\
				\hline
				$S^-$ &  lowering  operator for the atom\\
				\hline
				$g$ & coupling constant between the atom and the cavity field  \\
				\hline
				$\epsilon$ & coupling constant between the atom and the classical field\\
				\hline
				$\omega_0$& frequency for atomic transition\\
				\hline
				$\omega_a$& frequency for cavity mode\\
				\hline
				$\omega$  &  frequency for classical field.\\
				\hline
				$\rho$  &  density operator\\
				\hline
				$\ket{\psi}$  &  state vector\\
			\end{tabular}
		\end{ruledtabular}
	\end{table}
	\section{Photon number distribution}
	Photon number distribution (PND) is the probability distribution of finding $n$ photons in a given cavity field state and it can be obtained as the expectation of the density field in Fock state $|n\rangle$ as
	\begin{eqnarray}	
		\label{eq4}
		p(n) = \langle n|\rho|n\rangle
	\end{eqnarray}
	Now
	\begin{eqnarray}\nonumber
		\label{eq5}
		{\rho}_f & = & |\psi \rangle_f\,{}_f\langle \psi|\\
		& = & \sum_{n} \Big[|c_{a,n}|^2|n\rangle \langle n|+ c_{a,n} c_{b,n+1}^*|n\rangle \langle n+1|\\\nonumber
		& &  + c_{a,n}^* c_{b,n+1} |n+1\rangle \langle n| +
		|c_{b,n+1}|^2 |n+1\rangle \langle n+1|\Big]
	\end{eqnarray}
	Substituting \eqref{eq5} into \eqref{eq4}, the photon number distribution for the cavity field is obtained as
	\begin{eqnarray}\nonumber
		\label{eq6}
		p(n) & = & |c_n(0)|^2\left[{\cos}^2\left(\frac{\Omega_n t}{2}\right)+\frac{\delta^2}{{\Omega_n}^2} {\sin}^2\left( \frac{\Omega_n t}{2}\right)+\frac{4g^2(n+1)}{{\Omega_n}^2}{\sin}^2\left( \frac{\Omega_n t}{2}\right)\right]\\
		& = & |c_n(0)|^2
	\end{eqnarray}
	Here \eqref{eq6} gives the final expression of PND, where $c_n(0)$ is the probability
amplitude of the field alone.
	\begin{figure}[h]
		\centering
		\includegraphics[width=0.4\textwidth]{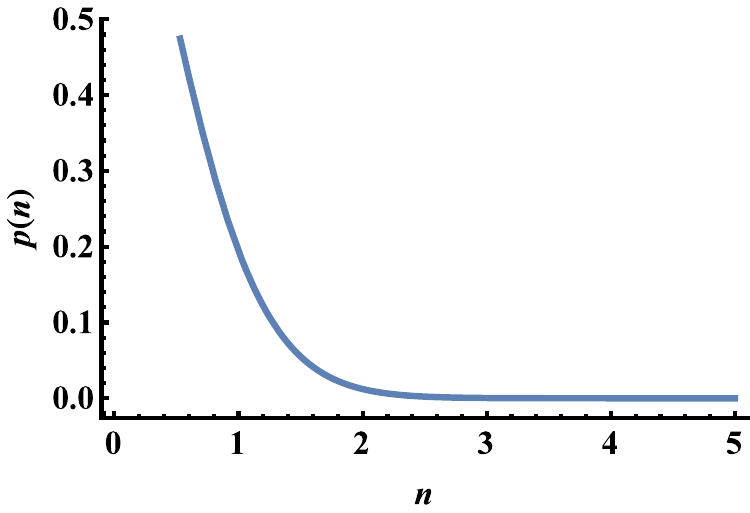}
		\caption{Photon number distribution of cavity field state is plotted against $n$ for $|\alpha|=0.5$.}
	\label{fig1}
\end{figure}
In Fig.~\ref{fig1}, we see the regular changes in PND as we are changing photon number $n$. It is observed that with increasing $n$, PND decreases exponentially.

\section{Wigner function}

The nonclassicality of a quantum state can be studied
in terms of its phase-space distribution characterized by
the Wigner distribution. For a quantum state $\rho$, the
Wigner function of the system is defined in terms of the
coherent state basis \cite{scully97,pathak,chatterjee} as
\begin{eqnarray*}
W(\beta, \beta^*) = \frac{2}{\pi^2}e^{2|\beta|^2} \int d^2\gamma{\langle-\gamma|\rho|\gamma\rangle e^{-2(\beta^*\gamma-\beta\gamma^*)}},
\end{eqnarray*}
$|\gamma\rangle$ is a coherent state.
By using the relation \cite{abramowitz72}
\begin{eqnarray*}
\sum_{n=k}^\infty n_{C_k}\,y^{n-k} = (1-y)^{-k-1},
\end{eqnarray*}
the Wigner function can be expressed in series form as \cite{moyacessa93}
\begin{eqnarray}
\label{eqn8}
W(\beta, \beta^*) = \frac{2}{\pi} \sum_{k=0}^\infty \left[(-1)^k \langle \beta,k|\rho|\beta,k\rangle\right] ,
\end{eqnarray}
where $|\beta,k\rangle$ is defined as the displaced Fock state. The
partial negative value of the Wigner function is a one
sided condition for the nonclassicality of the related state
\cite{wangz}, in the sense that one cannot conclude that the state
is classical when the Wigner function is positive everywhere. For example, the Wigner function of the squeezed
state is Gaussian and positive everywhere but it is a well-known nonclassical state. For a classical state, the positivity of the
Wigner function is a necessary condition but not a sufficient one. Hence, a state with a negative region
in the phase-space distribution is essentially nonclassical.

Now the displaced number state $|\beta, k\rangle$ can be expressed
in number state basis as following
\begin{eqnarray}\nonumber
\label{eqn9}
|\beta, k\rangle & = &
D(\beta)| k \rangle\\\nonumber
& = & e^{-\frac{|\beta|^2}{2}}\sum_{m=0}^{k} \frac{ \beta^{*{m}}}{m!} \sqrt{\frac{k!}{(k-m)!}}\sum_{p=0}^{\infty} \frac{ \beta^p}{p!}\sqrt{\frac{(k-m+p)!}{(k-m)!}}|k-m+p \rangle
\end{eqnarray}
Thus
\begin{eqnarray}\nonumber
\langle \beta,k|n \rangle & = &
e^{-\frac{|\beta|^2}{2}}\sum_{m=0}^{k} \frac{(-\beta)^{m}}{m!} \sqrt{\frac{k!}{(k-m)!}} \sum_{p=0}^{\infty}\frac{ \beta^{*{p}}}{p!}  \sqrt{\frac{(k-m+p)!}{(k-m)!}}\langle k-m+p|n\rangle\\\nonumber
& = & e^{-\frac{|\beta|^2}{2}} \sqrt{\frac{k!}{n!}} (\beta^*)^{n-k}L_{k}^{(n-k)}\left(|\beta|^2 \right),
\end{eqnarray}
where $L_{m}^{(k)}(x)=\sum_{n=0}^{m}  \frac{(-x)^{n} (m+k)!}{ n!(m-n)!(k+n)!}$ is the associated Laguerre polynomial \cite{agg}. Substituting \eqref{eq5} into \eqref{eqn8} we get
\begin{eqnarray}\nonumber
W(\beta, \beta^*) & = & \frac{2}{\pi} \sum_{k=0}^{\infty} (-1)^k \bigg[c_{a,n}c_{a,n}^* \langle\beta,k|n\rangle\langle n|\beta,k \rangle+ c_{a,n}c_{b,n+1}^* \langle \beta,k|n \rangle \langle n+1|\beta,k \rangle\\\nonumber
& & + c_{a,n}^*c_{b,n+1} \langle\beta,k|n+1\rangle \langle n|\alpha,k\rangle +c_{b,n+1}c_{b,n+1}^* \langle \alpha,k|n+1 \rangle \langle n+1|\alpha,k\rangle\bigg]\\\nonumber
& = & \frac{2e^{-|\beta|^2}}{\pi} \sum_{k=0}^{\infty} (-1)^k k! \bigg[|c_{a,n}|^2 \frac{|\beta|^{2(n-k)}}{n!} \left\{L_{k}^{(n-k)} \left(|\beta|^2 \right) \right\}^2\\
& & +|c_{b,n+1}|^2  \frac{|\beta|^{2(n+1-k)}}{(n+1)!} \left\{L_{k}^{(n+1-k)} \left(|\beta|^2 \right) \right\}^2\bigg]
\end{eqnarray}
	\begin{figure}[h]
		\centering
		\includegraphics[width=0.4\textwidth]{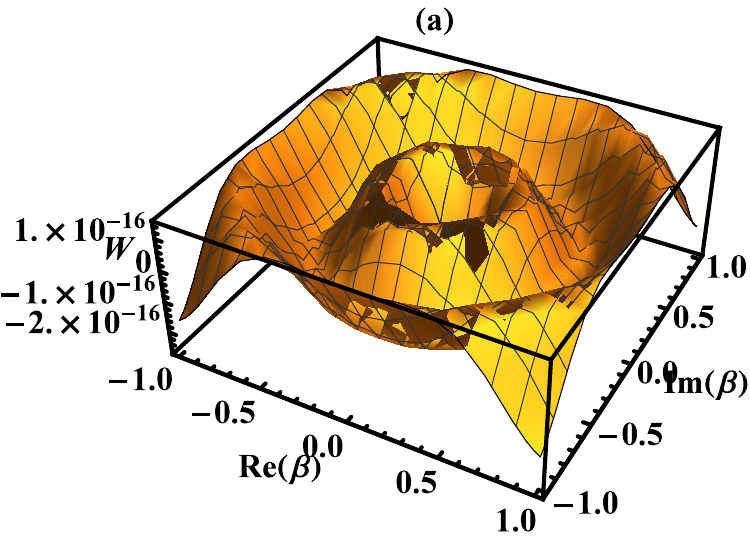}
		\includegraphics[width=0.4\textwidth]{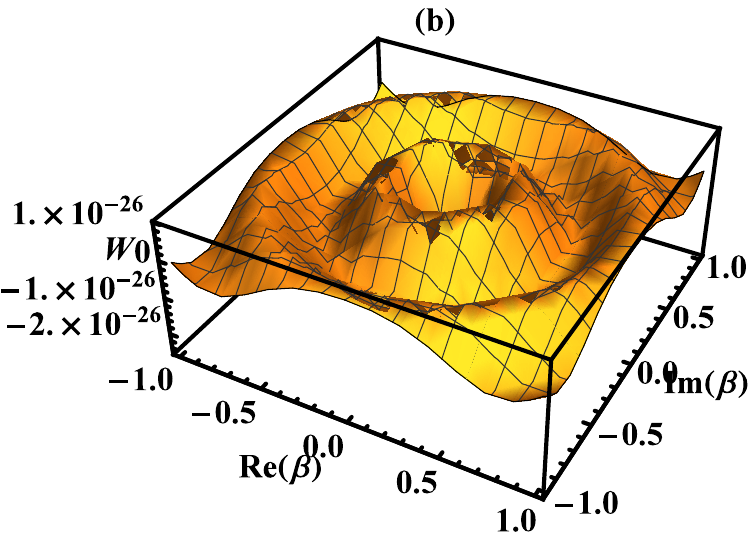}
		\includegraphics[width=0.4\textwidth]{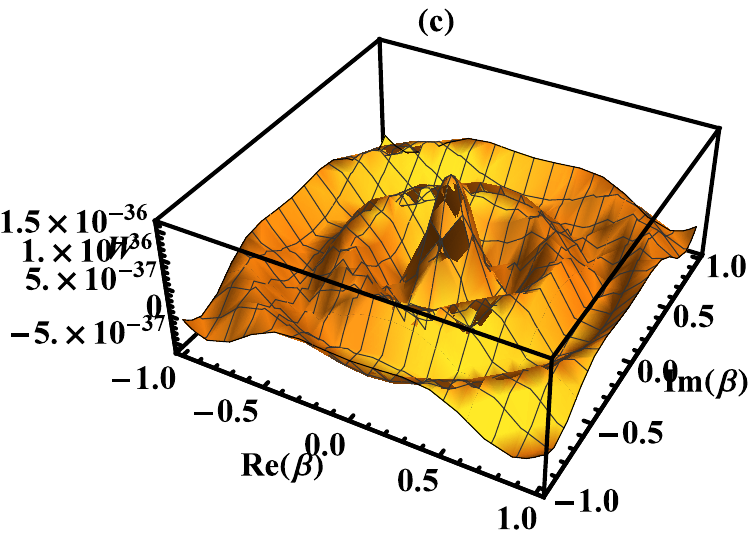}
		\caption{ Wigner distribution of cavity field state as a function of Real part of ($\beta$) and Imaginary part of ($\beta$) for fixed $|\alpha|=0.02$ and (a) $n=4$, (b) $n=7$, (c) $n=10$.}
		\label{fig2}
	\end{figure}
In Fig.~\ref{fig2}, the Wigner distributions for a fixed value of $|\alpha|$ and various distinct values of $n$ are shown. The cavity field exhibits nonclassical characteristics. When the number of photons is increased, keeping other parameters fixed, the Wigner function becomes less negative. As a result, a cavity field with more photons can prepare a classical Gaussian-like state.
\section{Mandel's $Q$ parameter }
Next to determine the photon statistics of a single-mode radiation field we consider the Mandel’s $Q$ parameter defined by \cite{mandel}
\begin{equation}
\label{eq14}
Q=\frac{\langle a^{{\dagger}^2} a^2 \rangle-(\langle a^{\dagger}a\rangle)^2} {\langle a^{\dagger}a\rangle}
\end{equation}
which measures the deviation of the variance of the photon number distribution of the considered state from the
Poissonian distribution of the coherent state. If $Q=0$, the distribution is Poissonian while for $-1\leq Q<0$ ($Q>0$), the field obeys super-Poissonian (sub-Poissonian) photon statistics. However, the lack of positivity in $Q$ is a sufficient condition for separating quantum states into nonclassical and classical regimes. For example, a state may be nonclassical even if $Q$ is positive \cite{agg}. The expectation values can be found as following
\begin{eqnarray*}
\langle a^{{\dagger}^2} a^2 \rangle & = & {}_f\langle \psi|a^{{\dagger}^2} a^2|\psi \rangle_f\\
& = & n(n-1)|c_{a,n}|^2+(n+1)n|c_{b,n+1}|^2 \\
\langle a^{\dagger}a\rangle & = & {}_f\langle \psi|a^{\dagger} a|\psi \rangle_f\\
& = & n|c_{a,n}|^2+(n+1)|c_{b,n+1}|^2
\end{eqnarray*}
Now substituting these values in \eqref{eq14}, we get the expression for  $Q$ for the cavity-driven field as
\begin{eqnarray}\nonumber
Q & = & \frac{n(n-1)|c_{a,n}|^2+(n+1)n|c_{b,n+1}|^2}{n|c_{a,n}|^2+(n+1)|c_{b,n+1}|^2}-\left(n|c_{a,n}|^2+(n+1)|c_{b,n+1}|^2\right)
\end{eqnarray}

	\begin{figure}[h]
		\centering
		\includegraphics[width=0.4\textwidth]{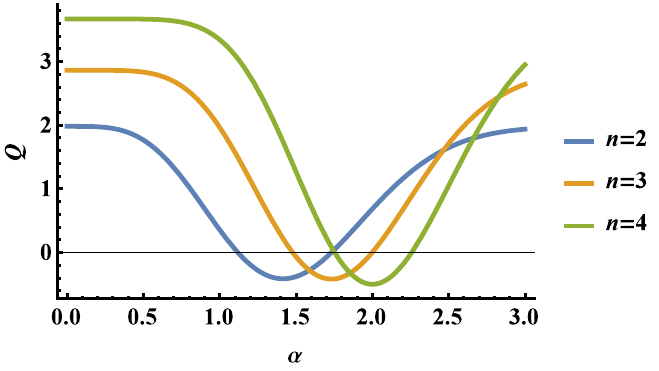}
		\caption{Mandel’s $Q$ parameter for a cavity field
state as a function of $\alpha$ ($\alpha$ assumed to be real) with different
photon numbers.}
		\label{fig3}
	\end{figure}
In order to see the variation of the Mandel’s $Q$ parameter against the coherent state amplitude $\alpha$, it is plotted
in Fig.~\ref{fig3}.
Here $Q$ curves go below zero for a certain limit of $\alpha$ which depicts that the cavity field enjoys sub-Poissonian character in a particular interval. It is observed that $Q$ usually exhibits super-Poissonian nature over the range of $\alpha$ and as $n$ increases, Mandel's $Q$ becomes more super-Poissonian.
\section{Squeezing properties}
In order to analyze the quantum fluctuations of the
field quadratures, we consider two Hermitian operators which are combinations of photon creation and annihilation operators as
$$\hat{x}=\frac{{a}+{a}^{\dagger}}{2},\,\,\,\,\,\,\hat{p}=\frac{{a}-{a}^{\dagger}}{2i}$$\\
with $\left[\hat{x},\hat{p}\right] = i/2 $. They follow the Heisenberg uncertainty principle expressed as
$\langle(\Delta\hat{x})^2\rangle\langle(\Delta\hat{p})^2\rangle\geq\frac{1}{16}$, and thus the quadrature squeezing occurs whenever $\langle(\Delta\hat{x})^2\rangle<\frac{1}{4}$ or $\langle(\Delta\hat{p})^2\rangle<\frac{1}{4}$. The squeezing parameters $s_x$ and $s_p$ are introduced as \cite{a10}
\begin{eqnarray}\nonumber
s_x & = & 4\langle(\Delta\hat{x})^2\rangle-1\\\nonumber
& = & 2\langle{a}^\dag{a}\rangle+\langle{a}^2\rangle+\langle{a}^{\dag2}\rangle-
\langle{a}\rangle^2-\langle{a}^\dag\rangle^2-2\langle{a}\rangle\langle{a}^\dag\rangle,
\end{eqnarray}
and
\begin{eqnarray}\nonumber
s_p & = & 4\langle(\Delta\hat{p})^2\rangle-1\\\nonumber
& = & 2\langle{a}^\dag{a}\rangle-\langle{a}^2\rangle-\langle{a}^{\dag2}\rangle+
\langle{a}\rangle^2+\langle{a}^\dag\rangle^2-2\langle{a}\rangle\langle{a}^\dag\rangle,
\end{eqnarray}
Squeezing occur in $\hat{x}$ or $\hat{p}$ quadrature if $-1<s_x<0$ or $-1<s_p<0$, respectively. The expectations can be evaluated as
\begin{eqnarray*}
\langle{a}^\dag{a}\rangle & = & n |c_{a,n}|^2+(n+1)|c_{b,n+1}|^2\\
\langle {a}^2\rangle & = & 0\\
\langle {a}^{\dagger^2}\rangle & = & 0\\
\langle {a}\rangle & = & \sqrt{n+1} c_{a,n}^*c_{b,n+1} \\
\langle {a}^\dagger \rangle & = & \sqrt{n+1} c_{a,n} c_{b,n+1}^*
\end{eqnarray*}
Substituting these values, the expressions for the squeezing parameters can be found.


	\begin{figure}[h]
		\centering
	\includegraphics[width=0.4\textwidth]{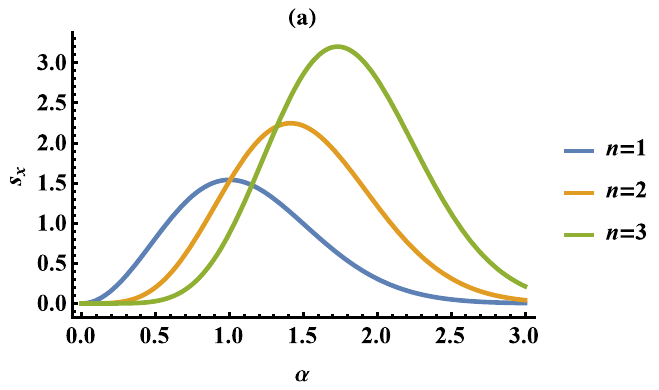}
		\includegraphics[width=0.4\textwidth]{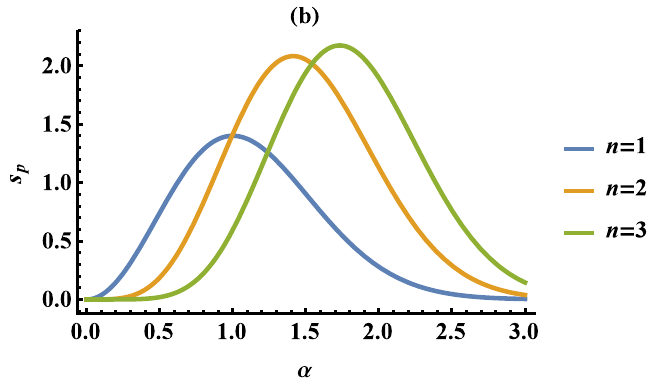}
			\caption{Plots of squeezing parameters $s_x$, $s_p$ against $\alpha$.}
		\label{fig4}
	\end{figure}
The negativity of squeezing is a sufficient criterion for identifying nonclassical quantum state regimes, rather than a necessary one. In fig.~\ref{fig4}, both the parameters $s_x$, $s_p$ are plotted against $\alpha$ and for different $n=1, 2$ and $3$. There is no squeezing in any of the quadratures, showing that the parameter $s_x$, $s_p$ are not so useful for identifying nonclassicality in the resulting cavity field.

\section{Conclusion}
In the article, we have obtained a cavity field state. The atom-cavity field is interacting in presence of a driven weak classical field. Assuming  the field is initially in a coherent state and the atom enters the cavity in its excited state $|a\rangle$, the cavity field state is obtained. The statistical features, namely the photon number distribution, Wigner function, Mandel's $Q$, and squeezing parameters, are calculated. The nonclassicality of the state has been depicted by two of them; the Wigner function's negativity and the Poissonian statistics of Mandel's $Q$ parameter. We have observed that the Wigner function has a partially negative region in phase space, indicating the nonclassicality of the cavity field state. Also, the $Q$ is negative in a certain range of $\alpha$.
We have also seen that the $Q$ function mostly displays super-Poissonian features. Here the squeezing parameters fail to depict the nonclassical behavior of the cavity field.

We can conclude that the cavity field state is showing nonclassical nature. The study of nonclassical states gives us the basic knowledge of quantum fluctuations and leads to more efficient way for new communication strategies. These generated states are widely used for testing fundamental quantum features such as teleportation and cryptography. These states are also useful in quantum computing and have many applications in quantum information processing.

\section{Acknowledgements}
Naveen's work is supported by the Council of Scientific and Industrial Research, Govt. of India (Grant no. 09/1256(0004)/2019-EMR-I).

\end{document}